# The infinite square well in a reformulation of quantum mechanics without potential function

A.D. Alhaidari [a], T.J. Taiwo [b]

[a] *Saudi Center for Theoretical Physics, P.O Box 32741, Jeddah 21438, Saudi Arabia.*
[b] *Department of Mathematics, University of Benin, Benin City, Edo State 300283, Nigeria.*

**Abstract**: Using a recent reformulation of quantum mechanics where the potential function is not required, we are able to obtain the energy spectrum and wavefunction associated with the infinite square well analytically. Therefore, this work constitutes an example of how to establish the correspondence between this approach and the standard formulation of quantum mechanics.

## 1. Introduction

Recently, Alhaidari et al. [1-3] proposed a reformulation of quantum mechanics where a potential function is not required. The objective was to find new analytically realizable quantum systems without the need for a prior knowledge of an analytic potential function. The hope was that by removing this constraint the class of exactly solvable systems might be enlarged beyond the conventional class to include systems whose potential functions cannot be realized analytically or those whose wave equation might be higher than second order, etc. The role of the potential function in the reformulation is replaced by orthogonal polynomials in the energy. The approach could be summarized as follows. We write the wavefunction as an infinite bounded sum over a complete set of square integrable functions as basis in configuration space. That is,

$$\psi(E, x) = \sum_n f_n^\mu(E)\phi_n(x) \tag{1}$$

where $\{\phi_n(x)\}_{n=0}^\infty$ is the $L^2$ basis functions in configuration space with coordinate $x$, $\{f_n^\mu(E)\}_{n=0}^\infty$ are the expansion coefficients at the energy $E$, and $\mu$ stands for a set of real parameter associated with the particular physical system. The basis set $\{\phi_n(x)\}_{n=0}^\infty$ satisfy the boundary conditions and contain only kinematical information such as the angular momentum, length scale, etc. On the other hand, structural and dynamical information about the system under study is contained only in the expansion coefficients, which we write as $f_n^\mu(E) = f_0^\mu(E) P_n^\mu(y)$, where $y$ is some proper function of the energy. Thus, $P_0^\mu(E) = 1$ and the completeness of the basis elements leads to the orthogonality relation [2]

$$\int \rho^\mu(y) P_n^\mu(y) P_m^\mu(y) dy = \delta_{nm}, \tag{2}$$

where $\rho^\mu(y) = [f_0^\mu(E)]^2$. Therefore, $\{P_n^\mu(y)\}_{n=0}^\infty$ becomes a set of polynomials in the energy variable $y$ that is orthogonal in some appropriate energy domain and with the weight function $\rho^\mu(y)$.

However, we also require that the physically relevant of these polynomials are those with the following asymptotics $(n \to \infty)$ form

$$P_n^\mu(y) \approx n^{-\tau} A^\mu(y) \times \cos\left[n^\xi \theta(y) + \delta^\mu(y)\right], \tag{3}$$

where $\tau$ and $\xi$ are real positive constants that depend on the particular energy polynomial. $A^\mu(y)$ is the scattering amplitude and $\delta^\mu(y)$ is the phase shift and both depend on energy and the physical parameters $\mu$. Bound states, if they exit, occur at the energies $\{E_n\}$ that make the scattering amplitude vanish. That is, when $A^\mu(y_m) = 0$. These bound states are either finite or infinite in number and we write the $m^{th}$ bound state as

$$\psi(E_m, x) = \sqrt{\omega^\mu(y_m)} \sum_n Q_n^\mu(y_m) \phi_n(x), \tag{4}$$

where $\{Q_n^\mu(y_m)\}$ are the discrete version of the orthogonal polynomials $\{P_n^\mu(y)\}$ and $\omega^\mu(y_m)$ is the associated discrete weight function. Therefore, in the absence of a potential function, the physical properties of the system in this formulation are deduced from the features of the orthogonal polynomials $\{P_n^\mu(y), Q_n^\mu(y_m)\}$ which include, the nature of the generating function, properties of the weight function, distribution and density of the polynomial zeros, associated three-term recursion relation, and so on.

In this paper, we like to illustrate the reformulation by giving an example of a simple system that corresponds to the infinite potential well. We show how to write the wavefunction and obtain the energy spectrum of the system. The energy polynomials of this system turns out to be the two parameters Meixner-Pollaczek polynomial whose asymptotics is in the standard form (3) giving the energy spectrum. This reformulation of quantum mechanics and the associated "*Tridiagonal Representation Approach* (TRA)" [4] enabled us to obtain exact solutions of well-known potentials in quantum mechanics and more.

## 2. The correspondence

The problem that we are addressing here is in one dimension and confined to a finite segment of length *a* with coordinate *x* such that $0 \leq x \leq a$. A complete set of square integrable basis functions that vanish on the boundaries $x = 0$ and $x = a$ could be chosen as follows

$$\phi_n(x) = \sqrt{2/a} \, \sin(n\pi x/a). \tag{5}$$

where $n = 0, 1, 2, \ldots$. Now, we choose the Meixner-Pollaczek polynomial as expansion coefficients in writing the wavefunction (1) as follows

$$\psi(E, x) = \sqrt{\rho^\mu(y;\theta)} \sum_n P_n^\mu(y;\theta) \phi_n(x), \tag{6}$$

where these polynomials and their weight function are shown in the Appendix as (A1) and (A2), respectively. Moreover, we choose the energy variable $y = ak/\pi$, where *k* is the wave number (i.e., linear momentum with $E = \tfrac{1}{2} k^2$). On the other hand, we choose the parameters $\mu \to 0$ and since $0 < \theta < \pi$ then we can take $\cos\theta = \frac{y^2 - 1}{y^2 + 1}$ assuming *y* to be real. However, since the system is physically confined in real space to the interval $0 \leq x \leq a$ then we expect that it does not have continuous energy states (scattering states) but only bound states. Thus, in the spectrum formula (A16) of the polynomial we replace $y \to iy$ giving $y_m^2 = +m^2$ and the wavefunction must be written in terms of the Meixner polynomials which are discrete version of the Meixner-Pollaczek polynomials. Consequently, the energy spectrum is now obtained from the spectrum formula as

$$E_m = \tfrac{1}{2} k_m^2 = \tfrac{1}{2}(\pi m/a)^2, \tag{7}$$

where $m = 0, 1, 2, \ldots$. This is identical to the energy spectrum of the one-dimensional infinite square potential well of width *a* [5]. On the other hand, if we leave $\mu$ arbitrary and choose the energy variable $y = \pi/ak$, then the energy spectrum becomes

$$E_m = \tfrac{1}{2} k_m^2 = \tfrac{1}{2}\left(\tfrac{\pi/a}{m+\mu}\right)^2. \tag{8}$$

This is a two-parameter spectrum whose corresponding potential box is not known in the standard formulation of quantum mechanics or cannot be realized analytically. That is, we cannot find a function $U(x)$ in

$$V(x) = \begin{cases} U(x) & , |x| < \tfrac{a}{2} \\ \infty & , x = \pm\tfrac{a}{2} \end{cases}. \tag{9}$$

such that the energy spectrum (8) solves the Schrödinger wave equation $\left[-\tfrac{1}{2}\tfrac{d^2}{dx^2} + V(x)\right]\psi_m(x) = E_m \psi_m(x)$.

## 3. Conclusion

In this note, we have given an example to show that our reformulation of quantum mechanics without a potential function could reproduce known physical systems. However, we did also demonstrate that this reformulation contains a new class of quantum mechanical systems that are not reproducible in the standard formulation.

## Appendix: Asymptotics of the Meixner-Pollaczek Polynomial

The normalized version of the Meixner-Pollaczek polynomials reads as follows

$$P_n^\mu(y, \theta) = \sqrt{\tfrac{\Gamma(n+2\mu)}{\Gamma(2\mu)\Gamma(n+1)}}\, e^{in\theta}\, {}_2F_1\!\left({-n, \mu+iy \atop 2\mu} \Big| 1 - e^{-2i\theta}\right), \tag{A1}$$

where $y \in [-\infty, +\infty]$, $\mu \geq 0$ and $0 < \theta < \pi$. Its normalized weight function is

$$\rho^\mu(y; \theta) = \tfrac{1}{2\pi\Gamma(2\mu)}(2\sin\theta)^{2\mu} e^{(2\theta-\pi)y} |\Gamma(\mu+iy)|^2, \tag{A2}$$

which means that $\int_{-\infty}^{+\infty} \rho^\mu(y;\theta) P_n^\mu(y,\theta) P_m^\mu(y,\theta) dy = \delta_{nm}$. To derive the large $n$ asymptotic formula for this polynomial, we apply the Darboux's method [6] to its generating function

$$\sum_{n=0}^\infty \tilde{P}_n^\mu(y,\theta) t^n = \left(1 - te^{i\theta}\right)^{-\mu+iy} \left(1 - te^{-i\theta}\right)^{-\mu-iy}, \tag{A3}$$

where $\tilde{P}_n^\mu(y,\theta)$ is defined as

$$\tilde{P}_n^\mu(y,\theta) = \sqrt{\tfrac{\Gamma(n+2\mu)}{\Gamma(2\mu)\Gamma(n+1)}}\, P_n^\mu(y,\theta). \tag{A4}$$

The generating function (A3) shows that there are two singularities in the complex $t$-plane at $t = e^{\pm i\theta}$. Thus, the dominant terms in a comparison function is

$$\left(1 - e^{2i\theta}\right)^{-\mu+iy}\left(1 - te^{-i\theta}\right)^{-\mu-iy} + \left(1 - e^{-2i\theta}\right)^{-\mu-iy}\left(1 - te^{i\theta}\right)^{-\mu+iy}. \tag{A5}$$

The expansion of one of the two $t$-dependent term is

$$\left(1-te^{-i\theta}\right)^{-\mu-iy} = \sum_{n=0}^{\infty} \frac{(\mu+iy)_n}{\Gamma(n+1)} e^{-in\theta} t^n . \tag{A6}$$

The other follows from this by the replacement $y \to -y$ and $\theta \to -\theta$. Therefore, the Darboux method states that

$$\tilde{P}_n^\mu(y;\theta) \approx \left(1-e^{2i\theta}\right)^{-\mu-iy} \frac{(\mu+iy)_n}{\Gamma(n+1)} e^{-in\theta} + \left(1-e^{-2i\theta}\right)^{-\mu-iy} \frac{(\mu-iy)_n}{\Gamma(n+1)} e^{in\theta}$$
$$= \frac{\left(1-e^{2i\theta}\right)^{-\mu-iy}}{\Gamma(\mu+iy)} \frac{\Gamma(n+\mu+iy)}{\Gamma(n+1)} e^{-in\theta} + \text{complex conjugate} \tag{A7}$$

where we have used $(z)_n = \frac{\Gamma(n+z)}{\Gamma(z)}$. If we also use the asymptotic form $\frac{\Gamma(n+a)}{\Gamma(n+b)} \approx n^{a-b}$ and $\Gamma(z) = |\Gamma(z)| e^{i \arg \Gamma(z)}$, this formula becomes

$$\tilde{P}_n^\mu(y;\theta) \approx \frac{\left(1-e^{2i\theta}\right)^{-\mu-iy}}{|\Gamma(\mu+iy)|} e^{-i\gamma} n^{\mu+iy-1} e^{-in\theta} + \text{complex conjugate} . \tag{A8}$$

where $\gamma = \arg \Gamma(\mu+iy)$. Now, we use $a^{ib} = e^{ib \ln a}$ to obtain

$$\tilde{P}_n^\mu(y;\theta) \approx \frac{n^{\mu-1}}{|\Gamma(\mu+iy)|} \left[ e^{iy \ln n} \; e^{-i\gamma} \; e^{-in\theta} \left(1-e^{2i\theta}\right)^{-\mu-iy} + \text{complex conjugate} \right] . \tag{A9}$$

Finally, we perform the following manipulation

$$\left(1-e^{2i\theta}\right)^{-\mu-iy} = \left[ e^{i\theta} \left( e^{-i\theta} - e^{i\theta} \right) \right]^{-\mu-iy} = e^{-\theta(y+i\mu)} \left(2 e^{-i\pi/2} \sin\theta \right)^{-\mu-iy}$$
$$= \frac{e^{(\pi/2-\theta)y}}{(2\sin\theta)^\mu} e^{-i\mu\theta} e^{i\mu\pi/2} (2\sin\theta)^{iy} = \frac{e^{(\pi/2-\theta)y}}{(2\sin\theta)^\mu} e^{-i\mu(\theta-\pi/2)} e^{iy \ln(2\sin\theta)} \tag{A10}$$

Thus, we can rewrite (A8) as follows

$$\tilde{P}_n^\mu(y;\theta) \approx \frac{n^{\mu-1}}{|\Gamma(\mu+iy)|} \frac{e^{(\pi/2-\theta)y}}{(2\sin\theta)^\mu} \left[ e^{iy \ln n} \; e^{-i\gamma} \; e^{-in\theta} \; e^{-i\mu(\theta-\pi/2)} \; e^{iy \ln(2\sin\theta)} + c.c. \right], \tag{A11}$$

which is equal to

$$\tilde{P}_n^\mu(y;\theta) \approx \frac{2 n^{\mu-1} e^{(\pi/2-\theta)y}}{(2\sin\theta)^\mu |\Gamma(\mu+iy)|} \cos\left[ (n+\mu)\theta + \gamma - y \ln(2n\sin\theta) - \mu(\pi/2) \right] \tag{A12}$$

Again, using $\frac{\Gamma(n+a)}{\Gamma(n+b)} \approx n^{a-b}$ and this result, we can write

$$P_n^\mu(y;\theta) \approx \frac{2 n^{-1/2} e^{(\pi/2-\theta)y}}{(2\sin\theta)^\mu |\Gamma(\mu+iy)|} \cos\left[ n\theta + \gamma - y \ln(2n\sin\theta) - \mu(\pi/2) \right] . \tag{A13}$$

This is in the standard format (3) with $\tau = \frac{1}{2}$ and $\xi = 1$. Moreover, Since $\ln n \approx o(n)$ as $n \to \infty$ then the $n$-dependent term $y \ln(2n \sin \theta)$ in the argument of the cosine in (A13) could be ignored relative to $n\theta$ giving the phase shift

$$\delta^\mu(y) = \arg \Gamma(\mu + iy) - \mu(\pi/2), \tag{A14}$$

and scattering amplitude

$$A^\mu(y) = \frac{2e^{(\pi/2-\theta)y}}{(2\sin\theta)^\mu |\Gamma(\mu+iy)|}. \tag{A15}$$

Therefore, bound states (if they exist) occur at energies where $A^\mu(y) = 0$, which means $\mu + iy = -m$ and $m$ is a non-negative integer which makes the gamma function in the denominator of (A15) blow up forcing the scattering amplitude to vanish. Thus, the infinite spectrum formula is

$$y^2 = -(m+\mu)^2, \quad m = 0, 1, 2, \ldots. \tag{A16}$$

The discrete version of these polynomials are the Meixner polynomials that are obtained from the continuous ones (A1) using the discrete condition $\mu + iy = -m$ and $\theta \to i\vartheta$ as

$$M_n^\mu(m;\vartheta) = \sqrt{\frac{\Gamma(n+2\mu)}{\Gamma(2\mu)\Gamma(n+1)}} e^{-n\vartheta} {}_2F_1\left(\begin{array}{c} -n,-m \\ 2\mu \end{array} \middle| 1 - e^{2\vartheta}\right), \tag{A17}$$

where $\vartheta \geq 0$. The corresponding normalized discrete weight function is

$$\omega_m^\mu(\vartheta) = (1-e^{-2\vartheta})^{2\mu} \frac{\Gamma(m+2\mu) e^{-2m\vartheta}}{\Gamma(2\mu)\Gamma(m+1)}, \tag{A18}$$

which means that $\sum_{m=0}^\infty \omega_m^\mu(\vartheta) M_n^\mu(m;\vartheta) M_{n'}^\alpha(m;\vartheta) = \delta_{n,n'}$.